\documentclass[%
 reprint,
superscriptaddress,
showpacs,preprintnumbers,
 amsmath,amssymb,
 aps,
prb,
]{revtex4-1}

\usepackage{graphicx}% Include figure files
\usepackage{dcolumn}% Align table columns on decimal point
\usepackage{bm}% bold math
\begin{document}

\preprint{APS/123-QED}

\title{Spin force and intrinsic spin Hall effect in spintronics systems}% Force line breaks with \\

\author{Cong Son Ho}
 %\email{sonhc@nus.edu.sg}
 \affiliation{
Department of Electrical and Computer Engineering, National University of Singapore,4 Engineering Drive 3, Singapore 117576, Singapore.
}
 \affiliation{
Data Storage Institute, Agency for Science, Technology and Research (A*STAR),
DSI Building, 5 Engineering Drive 1,Singapore 117608, Singapore.
}
\author{Mansoor B. A. Jalil}%
\affiliation{
Department of Electrical and Computer Engineering, National University of Singapore,4 Engineering Drive 3, Singapore 117576, Singapore.
}
\author{Seng Ghee Tan}
\affiliation{
Department of Electrical and Computer Engineering, National University of Singapore,4 Engineering Drive 3, Singapore 117576, Singapore.
}
\affiliation{
Data Storage Institute, Agency for Science, Technology and Research (A*STAR),
DSI Building, 5 Engineering Drive 1,Singapore 117608, Singapore.
}

\date{\today}% It is always \today, today,

\begin{abstract}
We investigate the spin Hall effect (SHE) in a wide class of
spin-orbit coupling systems by using spin force picture. We derive the general
relation equation between spin force and spin current and show that the
longitudinal force component can induce a spin Hall current, from which we
reproduce the spin Hall conductivity obtained previously using Kubo's formula.
This simple spin force picture gives a clear and intuitive explanation for SHE.

\end{abstract}

\pacs{71.70.Ej, 03.65.-w, 73.63.-b }% PACS, the Physics and Astronomy
\maketitle

%\tableofcontents

\section{INTRODUCTION}

Spin Hall effect (SHE) refers to the phenomenon in which a transverse pure spin
current is induced in response to a longitudinal applied electric field.\cite{Dya71,Hirsch99} The
generation of spin Hall current is associated with spin separation in the
transverse direction, which had been explained in many previous studies.\cite{Mura03,Sino04,Fu10}
Sinova \textit{et al.} derived a momentum-dependent spin polarization,\cite{Sino04} in
which electrons moving in the opposite transverse $(\pm{}y)$direction in a Rashba
system acquire opposite spin polarization, resulting in spin separation and SHE.
On the other hand, Murakami \textit{et al.} studied the effect in $p$-doped
semiconductors,\cite{Mura03} where the separation of electron spins is as a result of an
anomalous spin-dependent velocity. These two mechanisms were then brought under a
unified framework by Fujita \textit{et al. }invoking the gauge field in time
space.\cite{Fu10} In another work by Shen,\cite{Shen05} a heuristic picture of spin separation is
given in terms of the spin force. In this picture, electrons traveling in a 2DEG
system under the influence of a spin-orbit coupling (SOC) effect experience a
\textit{transverse} spin force, which induces separation of spin. However, the
spin separation due to transverse spin force was employed to describe the
\textit{Zitterbewegung} (jitter) motion of electrons, but not the SHE. In another study,\cite{Tan13}
 the spin force and spin Hall effect are shown to be linked. However, the spin Hall conductivity is still not derived.

Recently, we have also applied the spin force picture to study the SHE in
Rashba-Dresselhaus system\cite{Ho13} using non-Abelian gauge field and shown that the
\textit{longitudinal} spin force can induce a \textit{transverse} spin Hall
current, from which we recovered the universal spin Hall conductivities. This
picture is consistent with others,\cite{Mura03,Sino04,Fu10,Ada05} which assigned the underlying
mechanism of the SHE to the spin precession of electrons under acceleration.

In this paper, we generalize this spin force picture of the SHE for a general
SOC system, such as the cubic-Dresselhaus,\cite{Dress55} and heavy hole system based on
III-V semiconductor quantum wells.\cite{Gerchikov92} It may also be extended to other systems
governed by the same class of Hamiltonian involving the coupling between momentum
and a spin-like degree of freedom, such as the graphene systems.\cite{McCan06} We derived
the relation between the spin force and spin current, and showed that the
longitudinal force component is responsible for the SHC. From this general
relation, we recover the spin Hall conductivities obtained previously using
Kubo's formula. The spin force framework not only presents a unified picture of
SHE in a wide class of SOC systems, but also gives an intuitive picture of its
underlying mechanism, which is not obvious from the linear response or Kubo
theory.

\section{SPIN FORCE EQUATIONS}

\noindent \textit{Quantum spin force equation}\\

We begin with the general SOC Hamiltonian in presence of applied electric field:
\begin{equation}\label{eq1}
H=\frac{{\bf{p}}^2}{2m}+{\bf B}\left({\bf p}\right).\hat{\sigma}+ V(\bf{r})\ ,
%eq1
\end{equation}
where $m$ is the effective mass, $\bf{B}\left(\bf{p}\right)$is the momentum-dependent
effective magnetic field, which arises from the SOC effect, and $V({\bf{r}})=e{\bf E.r}$ with $\bf E$ is the applied electric field. The above Hamiltonian
has eigen-energies:
\begin{equation}\label{eq2}
{\epsilon{}}_{\pm{}}=\frac{p^2}{2m}\pm{}\left\vert{}B({\bf{p}})\right\vert{}
%eq2
\end{equation}
corresponding to eigen-vestors
\begin{equation}\label{eq2a}
\left\vert{\psi_{{\bf p},+}}\right\rangle{}=\left(\begin{array}{
cc}
 \cos{\frac{\Theta}{2}}e^{-i \Phi/2} \\
\\
\sin{\frac{\Theta}{2}}e^{i\Phi/2} \\
\end{array}\right), \left\vert{\psi_{{\bf p},-}}\right\rangle{}=\left(\begin{array}{
cc}
-\sin{\frac{\Theta}{2}}e^{-i \Phi/2} \\
\\
\cos{\frac{\Theta}{2}}e^{i\Phi/2} \\
\end{array}\right),
\end{equation}
respectively, where $\Theta$ and $\Phi$ are the spherical polar angles of the vector $\bf{B(p)}$ in $\bf{k}$-space.

The dynamics of electron in this system is described by equations of motion in
the Heisenberg picture:
\begin{eqnarray}
{\bf{v}}\equiv{}\dot{{\bf{r}}}&=&\frac{i}{\hbar{}}\left[H,{\bf{r}}\right]=\frac{{{\bf{p}}}}{m}+{\nabla{}}_{\bf{p}}\bf{B}.\hat{\sigma{}},\label{eq4}\\
\dot{\bf{p}}&=&\frac{i}{\hbar{}}\left[H,\bf{p}\right]=e\bf{E},\label{eq5}
\end{eqnarray}
where in \eqref{eq4}, we have made use of ${\bf{r}}=i\hbar{}{\nabla{}}_{\bf{p}}$. Likewise,
the spin dynamics can be shown to be governed by following equation:
\begin{equation}\label{eq6}
\frac{d\hat{\sigma{}}}{dt}=\frac{i}{\hbar{}}\left[H,\hat{\sigma{}}\right]=\frac{2}{\hbar{}}\left(\bf{B}\times{}\hat{\sigma{}}\right).
%eq5
\end{equation}

The force acting on electron can then be derived by taking time-derivative of
Eq. \eqref{eq4}, and using the results of Eq. \eqref{eq5} and \eqref{eq6} for the time derivatives of
$\bf{p}$ and $\hat{\sigma{}}$, respectively. This yields:
\begin{equation}\label{eq7}
\langle F_i\rangle=m\frac{d\langle v_i\rangle}{dt}=eE_i+m(U^0_{ij}+U^1_{ij})\langle{\sigma{}}_j\rangle
%eq6
\end{equation}
with
\begin{equation}\label{eq8}
U^0_{ij}=\frac{2}{\hbar{}}{\epsilon{}}_{jkl}\left({\nabla{}}_{p_i}B_k\right)B_l,\  U^1_{ij}=eE_k\frac{{\partial{}}^2B_j}{\partial{}p_i\partial{}p_k}.
%eq7
\end{equation}

\noindent In the above, we assume that repeated indices are summed up, and $\langle...\rangle$ denotes taking expectation value in spin-space.

We note that, in presence of an applied electric field, the linear response of the spin polarization can be written as
\begin{equation}\label{eq8a}
\langle\sigma_i\rangle=\langle\sigma_i^{0}\rangle+\langle\sigma_i^{1}\rangle,
\end{equation}
where $\sigma_i^{0}$ is the solution of equation \eqref{eq6} in the absence of electric field, and $\langle\sigma_i^{1}\rangle=A_{ij} E_j$ is the linear correction due to the electric field. It is obvious that the spin will alight along the effective SOC field when the electric field is absent, i.e., $\langle\sigma_i^{0}\rangle=\pm B_i/|B|$. To fulfill the normalization of the total spin polarization in Eq.\eqref{eq8a}, i.e., $\langle \sigma_i\rangle^2=1$, we must have
 \begin{equation}\label{eq8b}
\langle \sigma_i^1\rangle\langle \sigma_i^0\rangle=\pm\langle \sigma_i^1\rangle \frac{B_i}{|B|}=0, 
\end{equation}
which means that the electric field induces a spin correction that is perpendicular to the effective SOC field.

 With these, the force equation \eqref{eq7} can be rewritten as follow
\begin{equation}\label{eq7a}
\langle F_i\rangle=eE_i+mU^{0}_{ij}\langle{\sigma^1_j}\rangle+mU^{1}_{ij}\langle{\sigma^0_j}\rangle,
%eq6
\end{equation}
in which the higher order term in electric field is ignored, and $U^{0}_{ij}\langle{\sigma^0_j}\rangle={\epsilon{}}_{jkl}\left({\nabla{}}_{p_i}B_k\right)B_l B_j/|B|=0$ since ${\epsilon{}}_{jkl}$ is assymetric while $B_j B_l$ is symmetric in exchanging $(j,l)$.
\\

\noindent\textit{Classical spin force equation}\\

 Although in quantum mechanics, the force concept is not well-defined as a
consequence of the uncertainty principle, we can still establish a connection
between the expectation value of the force operator and the well-defined
classical force. While the former is derived from the Heisenberg's equation of
motion, the latter can be obtained from the energy of a physical system by
applying Hamilton's equations.  Indeed, if a physical system has energy
$\epsilon{}\left(\bf{p},\bf{r}\right),$which is a function of position and conjugate
momentum, its dynamics can be described by the coupled equations $\dot{\bf{r}}\
={\nabla{}}_{\bf{p}}\epsilon{}$ and $\bf{\dot{p}}=-{\nabla{}}_{\bf{r}}\epsilon{}$. Then,
the classical force acting on the system is given
by ${\bf{F}}^{cl}=m\ddot{\bf{r}}=m{({\bf \dot{p}}\nabla{}}_{\bf{p}}){\nabla{}}_{\bf{p}}\epsilon$.
 We now relate this classical force to the expectation value of force operator
$\left\langle{}{\bf{F}}\right\rangle{}$ of a quantum system, i.e.,
${\left\langle{}{\bf{F}}\right\rangle{}}{_{\psi_{{\bf p},n}}}={\bf{F}}^{cl}=m{({\bf \dot{p}}\nabla{}}_{\bf{p}}){\nabla{}}_{\bf{p}}\epsilon_n$,
with $n$ denoting the eigen-state index (we assume that the quantum system can
exist in different eigen-states $n$). In our present case, by considering the
eigen-energies in Eq. (2), the spin force is readily obtained as:

\begin{equation}\label{eq9}
{\left\langle{}F_i\right\rangle{}}_{\psi_{{\bf p},n}}^{cl}=eE_i+n\frac{{\partial{}}^2|B|}{\partial{}p_i\partial{}p_j}\
meE_j.
%eq8
\end{equation}
with $n=\pm$ being the eigen-branch index. In the above equation, the second term on the right hand side is odd either in $p_i$ or $p_j$ if $i\ne j$, and it is even when $j=i$; this means that upon averaging the above force equation over the Fermi sphere, only term with $j=i$ contributes to the total force. Therefore, if the electric field is just applied along the longitudinal $x$-direction, there only net longitudinal force exists in the system. 
Interestingly, we can express the above force as
\begin{equation}\label{eq9a}
{\left\langle{}F_i\right\rangle{}}_{\psi_{{\bf p},n}}^{cl}=eE_i+n\frac{\hbar^2}{4|B|^3} U_{ij}^0 U_{kj}^0 eE_k+nU_{ij}^1  \frac{B_j}{|B|}.
\end{equation}
%\noindent Equating the expectation value of the classical force with the force equation of Eq.\eqref{eq7a}, we have
%\begin{equation}\label{eq10}
%{\left\langle{}{\sigma{}}_j^1\right\rangle{}}_{{p,n}}=n\frac{\hbar^2}{4|B|^3}U_{kj}^0 eE_k.
%eq9
%\end{equation}

%\noindent This equation is consistent with previous results obtained by using time gauge field method\cite{Fu10}. The above spin correction also fulfills the normalization condition of the total spin as discussed in previous part. 
\noindent which bears a similarity to the spin force in Eq.\eqref{eq7a}.

Eqs. \eqref{eq7a} and \eqref{eq9} relates the spin polarization to the force (electric field) driving the electric current, and thus enables us to quantify other spin-dependent transport effects, e.g. spin Hall effect, or spin separation, in terms of the spin force. In following part, we will show that in general, the spin Hall current can be regarded as being induced by the spin force. For any general SOC system, we can thus derive spin Hall current and the associated spin Hall conductivity, once we have obtained the expression of spin force.

\section{SPIN CURRENT OPERATOR}

By definition, the spin current operator is $j_j^i=\frac{s}{2}\langle\left\{{\sigma{}}_i,v_j\right\}\rangle$ where $\{A,B\}$ denotes the
anti-commutation relation, $s$ is the spin of carriers (with
$s=\frac{\hbar{}}{2}$ for electron, and $s=\frac{3\hbar{}}{2}$ for heavy hole
in a Luttinger system). With the velocity operator given in Eq. (4), the spin
current operator then reads as
\begin{equation}\label{eq11}
\langle j_j^i\rangle=s\frac{p_j}{m}(\langle\sigma_i^1\rangle+\langle\sigma_i^0\rangle)-s\frac{\partial{}B_i}{\partial{}p_j}.
%eq10
\end{equation}
In the above, the first term depends on spin polarization which is induced by applied electric field, the second term represents the spin current in equilibrium state, i.e., in the absence of $\bf E$ field, while the last term relates to the variation of effective field in $k$-space. In our study, we focus on the spin Hall current contribution which is proportional to the electric field, i.e., the first term only:
\begin{equation}\label{eq12}
\langle j_j^i\rangle=s\frac{p_j}{m}\langle{\sigma{}}_i^1\rangle.
\end{equation}
The total spin Hall current $J^i_j$ can be obtained by integrating above expression over the momentum space. In the framework of linear response theory, the spin Hall current in semiconductors with SOC exhibits the general response of [3,4]
\begin{equation}\label{eq13}
J_j^i={\sigma{}}_{\textrm{sH}}{\epsilon{}}_{ijk}E_k,
%eq11
\end{equation}
where $\sigma_{\textrm{sH}}$ is the spin Hall conductivity. Thus, if an electric field is applied along one of the axes, e.g., the $x$-direction, there would be two non-zero transverse spin current components $j_y^z$ and $j_z^y$, which would in turn induce spin accumulation $\sigma^z$ and $\sigma^y$, respectively, and that $j_z^y=-j_y^z$. Moreover, there is only longitudinal spin force (along $x-$direction) acting on electron as discussed above. From now on, we will just consider this case for simplicity. With Eqs.\eqref{eq7a}, \eqref{eq12} and \eqref{eq13}, we can establish the relation between the longitudinal spin force acting on the electron and the resulting transverse spin current.

\begin{figure}
 \includegraphics[width=0.3\textwidth]{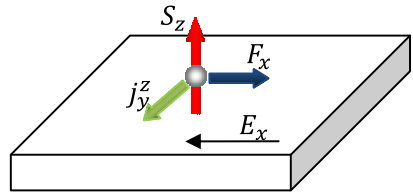}
 \caption{The correlation between vertical spin polarization, longitudinal spin force and spin Hall current. In a 3D system, there are two spin Hall current components $j^z_y$ and $j^y_z$ corresponding to the spin polarizations in $y-$ and $z-$direction, respectively.\label{Fig1}}
 \end{figure}

From Eq.\eqref{eq8b}, we have following identity: $\langle\sigma_x^{1}\rangle=-(B_y\langle\sigma_y^{1}\rangle+B_z\langle\sigma_z^{1}\rangle)/B_x$. With this, the longitudinal component of spin force (without the $eE_x$ term) in Eq.\eqref{eq7a} can be expressed as:
\begin{eqnarray}
\langle F_x\rangle=&&\frac{m^2}{s} \left(U^0_{xy}-U^0_{xx}\frac{B_y}{B_x}\right)\frac{\langle j^y_z\rangle}{p_z}\nonumber\\
&&+\frac{m^2}{s} \left(U^0_{xz}-U^0_{xx}\frac{B_z}{B_x}\right)\frac{\langle j^z_y\rangle}{p_y}\nonumber\\
&&+mU^1_{xj}\langle\sigma_j^0\rangle,\label{eq16a}
%eq12
\end{eqnarray}
where the spin polarizations have been replaced by the corresponding spin currents in Eq.\eqref{eq12}.
From Eq. \eqref{eq16a}, it can clearly be seen that the longitudinal spin force induces transverse Hall currents (Fig.\ref{Fig1}).By comparing the above spin force relations with the classical analogue [Eqs. \eqref{eq9} or \eqref{eq9a}], we can thus obtain the explicit expression for the spin Hall current, as well as the spin Hall conductivity.
Indeed, we can rewrite the force in Eq.\eqref{eq9a} as:
\begin{eqnarray}
{\left\langle{}F_x\right\rangle{}}_{\psi_{{\bf p},\pm}}^{cl}&&=\pm\frac{m\hbar^2eE_x}{4|B|^3} \left(U^0_{xy}-U^0_{xx}\frac{B_y}{B_x}\right) U_{xy}^0\nonumber\\
&&\pm\frac{m\hbar^2eE_x}{4|B|^3} \left(U^0_{xz}-U^0_{xx}\frac{B_z}{B_x}\right) U_{xz}^0\nonumber\\
&&\pm mU_{xj}^1  \frac{B_j}{|B|},\label{eq9b}
\end{eqnarray}
in which we have used the relation $U^{0}_{ij}B_j/|B|=0$.
\noindent By substituting above force expression into Eq.\eqref{eq16a}, the spin Hall current components are readily obtained as follows:
\begin{subequations}\label{eq17c}
\begin{eqnarray}
\langle j^y_z\rangle_{\psi_{{\bm p}\pm}}=\pm s\frac{\hbar^2eE_x}{4m|B|^3} p_z U_{xy}^0 \label{eq17a},\\
\langle j^z_y\rangle_{\psi_{{\bm p}\pm}}=\pm s\frac{\hbar^2eE_x}{4m|B|^3} p_y U_{xz}^0\label{eq17b}.
\end{eqnarray}
\end{subequations}

If the electron motion is confined to a 2D plane ($x-y$ plane), we have $B_z({\bf k})=0$, and $U^0_{xy}=0$ following Eq.\eqref{eq8}. This means that the transverse spin Hall current in 2D system is given by Eq.\eqref{eq17b}.

Eq.\eqref{eq16a} and Eqs.\eqref{eq17c} are our main results. By deriving the spin force in Heisenberg picture and its classical counterpart using Hamilton's equations, we have explicitly obtained the transverse spin Hall currents. In the next section, we will apply our analysis to a wide class of of systems for describing the SHE. 

\section{RESULTS AND DISCUSSIONS}

We will now illustrate the utility of the spin force picture in evaluating the spin Hall current in exemplary 2D and 3D SOC systems, the corresponding Hamiltonian of which is listed in Table \ref{tab1}, together with their respective effective magnetic fields $\bf{B} (\bf{p} )$. For each system, we will consider the spin force equation [Eqs. \eqref{eq16a}], and evaluate the $U_{ij}$ matrix based on Eq. \eqref{eq8}. Next, we calculate the classical force based on its expectation value when the electron is in the eigenstate $\psi_{{\bf{p}},n}$ [Eq. (9)]. By equating the classical and quantum mechanical spin force expression, we obtain the expression for the spin current for the state $\psi_{{\bf{p}},n}$ corresponding to the $n$-th eigen-branch and momentum. Finally, the total spin current of the system is obtained by summing over the momentum space and eigen-branches of the system.\\

\begin{table*}
\caption{\label{tab1}Various SOC systems and their corresponding Hamiltonian and effective SOC field. Here, $k_\pm=k_x\pm i k_y$, $\sigma_\pm=\sigma^x\pm i\sigma^y$, and c.p denotes cyclic permutation among (x,y,z).}
\begin{ruledtabular}
\begin{tabular}{ccc}
SOC system&Hamiltonian&$\bf{B(\bf{p})}$\\\hline
Rashba-Dresselhaus (2D)&${\bf{p}}^2/2m+\alpha{}\left(p_x{\sigma{}}^y-p_y{\sigma{}}^x\right)+\beta{}\left(p_y{\sigma{}}^y-p_x{\sigma{}}^x\right)$&$\left(\begin{array}{
ccc}
-\alpha{}p_y-\beta{}p_x \\
\alpha{}p_x+\beta{}p_y \\
0
\end{array}\right)$\\
Heavy hole in QW(2D)&${\bf{p}}^2/2m+i\lambda/2 (k_-^3 \sigma_+-k_+^3 \sigma_- )$&$\lambda\left(\begin{array}{
ccc}
3k_x^2k_y-k_y^3 \\
3k_xk_y^2-k_x^3 \\
0
\end{array}\right)$\\

$k^3$-Dresselhaus(3D)&${\bf{p}}^2/2m+\eta[k_x (k_y^2-k_z^2 ) \sigma^x+c.p]$&$\eta{}\left(\begin{array}{
ccc}
k_x\left(k_y^2-k_z^2\right) \\
k_y\left(k_z^2-k_x^2\right) \\
k_z\left(k_x^2-k_y^2\right)
\end{array}\right)$

\end{tabular}
\end{ruledtabular}
\end{table*}

\noindent\textit{a) Linear Rashba-Dresselhaus system}\\

From Eq. \eqref{eq16a}, the equation relating the spin force to the spin current in this
system is readily found to be :
\begin{equation}\label{eq18}
F_x=\frac{4m^2\left({\alpha{}}^2-{\beta{}}^2\right)}{{\hbar{}}^2}j_y^z.
%eq14
\end{equation}
Meanwhile, considering Eq. \eqref{eq9}, the classical force corresponding to
eigen-states${\psi{}}_{p,\pm{}}$is
\begin{equation}\label{eq19}
{\left\langle{}F_x\right\rangle{}}_{{\psi{}}_{p,\pm{}}}=\pm{}\frac{m{eE_x\left({\alpha{}}^2-{\beta{}}^2\right)}^2\sin{{\theta{}}^2}}{p{\left({\alpha{}}^2+{\beta{}}^2+2\alpha{}\beta{}\sin{2\theta{}}\right)}^{3/2}}
%eq15
\end{equation}
for the $\pm{}$ eigen-branches. It is obvious that the spin force in both Eqs.\eqref{eq18}
and \eqref{eq19} will vanish if $\alpha=\pm\beta{}$. For the case of $\alpha \ne\pm \beta{}$, by equating \eqref{eq18} and \eqref{eq19}, integrating over $\bf{p}$ and summing over the contribution of the two eigen-branches, the spin current is readily shown to be
\begin{equation}\label{eq20}
J_y^z\left(\textrm{sH}\right)=\frac{{\alpha{}}^2-{\beta{}}^2}{\left\vert{}{\alpha{}}^2-{\beta{}}^2\right\vert{}}\left(\frac{eE_x}{8\pi{}}\right)
%eq16
\end{equation}
a result which is consistent with previous calculations based on Kubo linear response theory.\cite{Sino04,Shen04}

It is instructive at this point to note that for the linear Rashba-Dresselhaus system, the quantum spin force operator in Eq. \eqref{eq16a}, obtained from the general form of Eq.~\eqref{eq13}, can be couched in terms of the Lorentz force in the non-Abelian gauge formalism. This may be seen by rewriting the Rashba-Dresselhaus Hamiltonian in the form of non-Abelian(or Yang-Mills) gauge fields as follows:
\begin{equation}\label{eq21}
H_{{\textrm RD}}^{\textrm{YM}}=\frac{{\left({\bf{p}}-e{\cal{A}}\right)}^2}{2m},
\end{equation}
where the non-Abelian gauge field is
\begin{equation}\label{eq22}
{\cal{A}}=\left({\cal{A}}_x,{\cal{A}}_y,0\right)=\frac{m}{e}\left(-\alpha{}{\sigma{}}^y+\beta{}{\sigma{}}^x,\alpha{}{\sigma{}}^x-\beta{}{\sigma{}}^y,0\right).
\end{equation}
Then, the effective Yang-Mills magnetic field associated with this gauge is given by
\begin{equation}\label{eq23}
{\cal{B}}^{\textrm{YM}}=B_z\hat{z}=-\frac{ie}{\hbar{}}\left[{\cal{A}}_x,{\cal{A}}_y\right]=\frac{2m^2\left({\alpha{}}^2-{\beta{}}^2\right)}{e\hbar{}}{\sigma{}}^z\hat{z}.
\end{equation}
The above field then exerts a Lorentz-like force on the
electron: ${\bm F}^{YM}=e\left({\bm v}\times{}{\cal{B}}_z^{YM}\hat{z}\right)$. Substituting the above
expression for the non-Abelian ${\cal{B}}_z^{YM}$, we have
\begin{equation}\label{eq24}
{\bm{F}}^{\textrm{YM}}=\frac{2m\left({\alpha{}}^2-{\beta{}}^2\right)}{e\hbar{}}\left(v\times{}\hat{z}\right){\sigma{}}^z.
\end{equation}

Considering the expression for the spin current operator in Eq. \eqref{eq11}, the above force equation can then be rewritten as
\begin{equation}\label{eq25}
F_i^{\textrm{YM}}={\epsilon{}}_{ijz}\frac{4m^2\left({\alpha{}}^2-{\beta{}}^2\right)}{{\hbar{}}^2}j_j^z,
\end{equation}
which is consistent with Eq. \eqref{eq18}.

\noindent \textit{b) Heavy hole quantum well system}\\

For heavy hole in quantum well, the spin of carriers is $s=3\hbar/2$, so that the
force equation \eqref{eq16a} is
\begin{equation}\label{eq26}
\langle F_x\rangle=\frac{4m^2{\lambda{}}^2p^4}{{\hbar{}}^8}\langle j_y^z\rangle \pm\frac{6\lambda^2 eE_x}{\hbar^6|B|}p^4\cos 2\theta,
%eq17
\end{equation}
in which the magnitude of the SOC field is $|B|=\frac{\lambda{}p^3}{{\hbar{}}^3}$, which also yields
the classical force
\begin{equation}\label{eq27}
{\left\langle{}F_x\right\rangle{}}_{\pm{}}=\pm{}\frac{3m\lambda{}peE_x}{2{\hbar{}}^3}(3+\cos{2\theta{}).\
}
%eq18
\end{equation}
From these two equations, the spin current reads
\begin{equation}\label{eq28}
j_y^z=\pm\frac{9eE_x{\hbar{}}^5}{4\lambda{}mp^3}\sin^2\theta,
%eq19
\end{equation}
which is summed over momentum space and two branches to give total value:
\begin{equation}\label{eq29}
J_y^z(\textrm{sH})=-\frac{9eE_x{\hbar{}}^3}{16\lambda{}m\pi{}}\left(\frac{1}{p_{F-}}-\frac{1}{p_{F+}}\right)
%eq20
\end{equation}
This result is consistent with previous findings obtained via the Kubo formula.\cite{Schli05}\\

\noindent \textit{c) Cubic $k^3$-Dresselhaus}\\

In $k^3$- Dresselhaus system, there are two spin Hall current components $j^z_y$ and $j^y_z$ given in Eqs.\eqref{eq17c}. Introducing the chiral spin current as $j_{\textrm{chir}}=(j^z_y-j^y_z)/2$, which explicitly reads:
\begin{eqnarray}
j_{\textrm{chir}}=\pm\frac{eE_x\eta^2p_x^2(p_y^2-p_z^2)^2}{4\hbar^4m|B|^3}.
\end{eqnarray}
The total spin Hall current is then:
\begin{equation}\label{eq32}
J_{\textrm{chir}}\left(\textrm{sH}\right)=\frac{eE_x\eta^2}{4\hbar^4m}\int_{p_-}^{p_+}\frac{d^3{\bm p}}{{\left(2\pi{}\hbar{}\right)}^3}\frac{p_x^2(p_y^2-p_z^2)^2}{|B|^3}
\end{equation}
Using the inter-band relation with small spin-split $(p_+-p_- )=-2m|B|/p_F$, with $p_F=(p_++p_-)/2$ is the average Fermi momentum, the above SHC is simplified to

\begin{equation}\label{eq33}
J_{\textrm{chir}}\left(\textrm{sH}\right)=eE_x\left(\frac{k_F}{12{\pi{}}^2}\right)
%eq23
\end{equation}
which recovers previous results.\cite{Bern04}
\\

In summary, we have described the spin Hall effect in various semiconductor SOC systems by invoking the spin force picture, both in the quantum mechanical and classical sense. The former relates the longitudinal force to a transverse spin current carrying a perpendicular spin polarization via the Heisenberg's equation of motion. For the specific case of linear Rashba-Dresselhaus system, the spin force can be related to the Lorentz-like force arising from a non-Abelian (Yang-Mills) field. The classical spin force equation then enables an explicit evaluation of the transverse spin current and hence the spin Hall conductivity. The calculated spin Hall conductivities are consistent with those obtained via other methods.

\begin{acknowledgments}
We gratefully acknowledge the SERC Grant No. 092 101 0060 (R-398-000-061-305) for financial support.
\end{acknowledgments}
%\bibliography{SpinForceSHE}
\providecommand{\noopsort}[1]{}\providecommand{\singleletter}[1]{#1}%
\end{document}